\def\CMP{\sevenrm Commun.\ Math.\ Phys.}

\def\JMP{\sevenrm J.\ Math.\ Phys.}
%
%
\def\today{\number\day .\space\ifcase\month\or
January\or February\or March\or April\or May\or June\or
July\or August\or September\or October\or November\or December\fi, \number \year}
%
%
\newcount \theoremnumber
\def\cleartheoremnumber{\theoremnumber = 0 \relax}

\newcount \subheadlinenumber

\def\SHL #1 {\goodbreak
            \cleartheoremnumber
            \vskip 1cm
            \advance \subheadlinenumber by 1
            {\fourteenrm \noindent {\the\headlinenumber}.{\the\subheadlinenumber}. #1}
            \nobreak \vskip.8cm \rm \noindent}

\def\Prop #1 {
             \advance \theoremnumber by 1
             \vskip .6cm 
             \goodbreak 
             \noindent
             {\bf Proposition {\the\headlinenumber}.{\the\theoremnumber}.}
             {\sl #1}  \goodbreak \vskip.8cm}

\def\Conj#1 {
             \advance \theoremnumber by 1
             \vskip .6cm  
             \goodbreak 
             \noindent
             {\bf Conjecture {\the\headlinenumber}.{\the\theoremnumber}.}
             {\sl #1}  \goodbreak \vskip.8cm} 

\def\Th#1 {
             \advance \theoremnumber by 1
             \vskip .6cm  
             \goodbreak 
             \noindent
             {\bf Theorem {\the\headlinenumber}.{\the\theoremnumber}.}
             {\sl #1}  \goodbreak \vskip.8cm}

\def\Lm#1 {
             \advance \theoremnumber by 1
             \vskip .6cm  
             \goodbreak 
             \noindent
             {\bf Lemma {\the\headlinenumber}.{\the\theoremnumber}.}
             {\sl #1}  \goodbreak \vskip.8cm}

\def\Cor#1 {
             \advance \theoremnumber by 1
             \vskip .6cm  
             \goodbreak 
             \noindent
             {\bf Corollary {\the\headlinenumber}.{\the\theoremnumber}.}
             {\sl #1}  \goodbreak \vskip.8cm} 
%
%
\newcount \equationnumber

\newcount \refnumber

\def\[]    {\global 
            \advance \refnumber by 1
            [{\the\refnumber}]}

\def\# #1  {\global 
            \advance \equationnumber by 1
            $$ #1 \eqno ({\the\equationnumber}) $$ }

\def\% #1 { \global
            \advance \equationnumber by 1
            $$ \displaylines{ #1 \hfill \llap ({\the\equationnumber}) \cr}$$} 

\def\& #1 { \global
            \advance \equationnumber by 1
            $$ \eqalignno{ #1 & ({\the\equationnumber}) \cr}$$}
%
%
\newcount \Refnumber

\def\Ref #1 #2 #3 #4 #5 #6  {\ninerm \global
                             \advance \Refnumber by 1
                             {\ninerm #1,} 
                             {\ninesl #2,} 
                             {\ninerm #3.} 
                             {\ninebf #4,} 
                             {\ninerm #5,} 
                             {\ninerm (#6)}\nobreak} 
\def\Bookk #1 #2 #3 #4       {\ninerm \global
                             \advance \Refnumber by 1
                             {\ninerm #1,}
                             {\ninesl #2,} 
                             {\ninerm #3,} 
                             {(#4)}}
\def\Book{\cr
{\the\Refnumber} &
\Bookk}
\def\Reff{\cr
{\the\Refnumber} &
\Ref}
\def\REF #1 #2 #3 #4 #5 #6 #7   {{\sevenbf [#1]}  & \hskip -9.5cm \vtop {
                                {\sevenrm #2,} 
                                {\sevenrm ``#3'',} 
                                {\sevenrm #4} 
                                {\sevenbf #5,} 
                                {\sevenrm #6} 
                                {\sevenrm (#7)}}\cr}
\def\BOOK #1 #2 #3 #4  #5   {{\sevenbf [#1]}  & \hskip -9.5cm \vtop {
                             {\sevenrm #2,}
                             {\sevenit #3,} 
                             {\sevenrm (#4,} 
                             {\sevenrm #5).}}\cr}
\def\HEP #1 #2 #3 #4     {{\sevenbf [#1]}  & \hskip -9.5cm \vtop {
                             {\sevenrm #2,}
                             {\sevenit #3,} 
                             {\sevenrm #4.}}\cr}
%
%
\def\bull{$\sqcup \kern -0.645em \sqcap$}
%
%
\def\Def#1{\vskip .3cm \goodbreak \noindent
                                     {\bf Definition.} #1 \goodbreak \vskip.4cm}
\def\Rem#1{\vskip .4cm \goodbreak \noindent
                                     {\it Remark.} #1 \goodbreak \vskip.5cm }

\def\Pr#1{\goodbreak \noindent {\it Proof.} #1 \hfill \bull  \goodbreak \vskip.5cm}

%
%
\def\*{\vskip 1.0cm}      

%
%
\newcount \headlinenumber

\newcount \headlinesubnumber
\def\clearheadlinesubnumber{\headlinesubnumber = 0 \relax}
\def\Hl #1 {\goodbreak
            \cleartheoremnumber
            \clearheadlinesubnumber
            \advance \headlinenumber by 1
            {\bf \noindent {\the\headlinenumber}. #1}
            \nobreak \vskip.4cm \rm \noindent}

\def\SHL #1 {\goodbreak
            \cleartheoremnumber
            \vskip 1cm
            \advance \subheadlinenumber by 1
            {\rm \noindent {\the\headlinenumber}.{\the\subheadlinenumber} #1}
            \nobreak \vskip.4cm \rm \noindent}

\font\twentyrm=cmr17
\font\fourteenrm=cmr10 at 14pt

\font\sevenit=cmti7

\font\css=cmss10
\font\Rosch=cmr10 at 9.85pt
\font\Cosch=cmss12 at 9.5pt
\font\rosch=cmr10 at 7.00pt
\font\cosch=cmss12 at 7.00pt
\font\nosch=cmr10 at 7.00pt
%
%
%
%
%
%
%
%
%
%
%
%
%
%
%
%
\def\Z                 {\hbox{{\css Z}  \kern -1.1em {\css Z} \kern -.2em }}
\def\R                 {\hbox{\raise .03ex \hbox{\Rosch I} \kern -.55em {\rm R}}}
\def\N                 {\hbox{\rm I \kern -.55em N}}
\def\C                 {\hbox{\kern .20em \raise .03ex \hbox{\Cosch I} \kern -.80em {\rm C}}}

\def\r                 {\hbox{\raise .03ex \hbox{\rosch I} \kern -.45em \hbox{\rosch R}}}
\def\n                 {\hbox{\hbox{\rosch I} \kern -.45em \hbox{\nosch N}}}
\def\c                 {\hbox{\raise .03ex \hbox{\cosch I} \kern -.70em \hbox{\rosch C}}}

\def\z                 {\hbox{\kern 0.2em {\cal z}  \kern -0.6em {\cal z} \kern -0.3em  }}
\def\1                 {\hbox{\rm \thinspace \thinspace \thinspace \thinspace
                                  \kern -.50em  l \kern -.85em 1}}
\def\unit                 {\hbox{\sevenrm \thinspace \thinspace \thinspace \thinspace
                                  \kern -.50em  l \kern -.85em 1}}
%
%
%
%
%
%
%
%
%
%
%
%
\def\Tr                {{\rm Tr}}
\def\A                 {{\cal A}}

\def\B                 {{\cal B}} 

\def\H                 {{\cal H}} 
 
\def\O                 {{\cal O}}

%

%
%
%
%
%
%
%


\nopagenumbers
\def\Draft  {\hbox{Preprint \today}}
\def\firstheadline{\hss \hfill  \Draft  \hss} 
\headline={
\ifnum\pageno=1 \firstheadline
\else 
\ifodd\pageno \rightheadline 
\else \leftheadline \fi \fi}
\def\rightheadline{\sevenrm Some Comments on Entanglement and Local Thermofield Theory
\hfill \folio } 
\def\leftheadline{\sevenrm \folio \hfill CHRISTIAN D.\ J\"AKEL}
\voffset=2\baselineskip
\magnification=\magstep1
%
%
%
%
%
%
%

\vskip 1cm

\noindent
{\twentyrm Some Comments on Entanglement and}

\vskip .5cm
\noindent
{\twentyrm Local Thermofield Theory}
                  
\vskip 1cm
\noindent
{\sevenrm Christian D.\ J\"akel

\noindent 
{\sevenit Insitut f\"ur theoretische Physik, Universit\"at Innsbruck, Austria}

\noindent 
{\sevenrm E-mail: christian.jaekel@uibk.ac.at} }



\vskip .5cm     
\noindent {\sevenrm We combine recent results of Clifton and Halvorson [1]
with structural results of the author [2--5] concerning the local observables in thermofield theory.
An number of interesting consequences are discussed.}

\vskip 1 cm

%
%
%
%
%

\Hl{INTRODUCTION}

\noindent
Entangled states are the starting point for quantum information theory, quantum cryptography and quantum
teleportation. 
As long as there are only a few observables experimentally accessible it can be rather
difficult to entangle a given system with respect to these observables\footnote{$^\dagger$}
{\sevenrm Quantum information theory is usually set up in a finite dimensional Hilbert space.
It is left up to the experimental physicists to realize an apparatus
which can approximately be described by such a theory (by neglecting the other degrees of freedom
of the system).}. If however,
the energy--momentum density distributions for bounded space--time regions are considered 
to be observable (as one usually does in quantum field theory), then entanglement is 
a generic feature (see e.g., [6][7]). As we will discuss in this short letter,
a thermal background sustains these effects: 

\vskip .3cm
\halign{ \indent #  \hfil & \vtop { \parindent = 0pt \hsize=34em
                            \strut # \strut} 
\cr 
$ \bullet$    & with respect to measurements in
spacelike separated space--time regions the set of entangled states is norm
dense (see Corollary 5.2 below) 
in the set of all states which only locally differ from the thermal background
(these states are described by density matrices in the Hilbert space $\H_\beta$
distinguished by the thermal state);
\cr
$ \bullet$   & any state, which is described by a density matrix acting on the
Hilbert space $\H_\beta$, can approximately (in the norm topology) be prepared  (see Theorem 3.1 below)
by a strictly local operation (i.e., an operation 
performed in an arbitrary bounded open space--time region $\O \subset \R^4$)
acting on the thermal equilibrium state;
\cr
$ \bullet$   & if $P$ and 
$Q$ are two nontrivial
projections (representing YES-NO experiments)
which are localized in spacelike separated space--time regions, 
then the product PQ can not vanish identically (see Theorem 4.1 below);
\cr
$ \bullet$ & Let $E$ represent a single YES-No experiment which can be performed in a bounded
space--time region $\O$. Given an arbitrary density matrix $\rho \in \B(\H_\beta)$
one can perform (see Theorem 4.3 below) an operation $V$ (more precisely, one can find an isometry)
in a slightly larger space--time region $\hat {\O}$
such that the density matrix $ \rho_V := V \rho V^*$ satisfies
\# { \Tr \, \rho_V E = 1  \quad \hbox{and} 
\quad \Tr \, \rho_V B  = \Tr \, \rho B \qquad 
\forall B \hbox{ observable in $\hat {\cal O}'$}.}
The state given by the density matrix $\rho$ remains
completely unchanged in the spacelike complement $\hat {\cal O}'$ of $\hat {\O}$.
\cr
$ \bullet$ &  for any pair $E,F$ of
nontrivial projections (i.e., YES-NO experiments)
which are localized in spacelike separated space--time regions 
and for all $\lambda, \mu \in [0, 1]$ 
there exists (see (25) below) a density matrix $\rho \in \B(\H_\beta)$
such that 
\# { \Tr \, \rho E = \lambda \quad \hbox{and}  \quad \Tr \, \rho F = \mu.} 
Moreover, for any pair of density matrices $\rho_1, \rho_2 \in \B(\H_\beta)$
and any pair of spacelike separated space--time regions $\O_1, \O_2$
there exists a density matrix $\rho$ such that
\# {\Tr \, \rho_1 A = \Tr \, \rho A  \qquad \forall A \hbox{ observable in $\O_1$}}
and 
\# {\Tr \, \rho_2 B = \Tr \, \rho B  \qquad \forall B  \hbox{ observable in $\O_2$} .} 
\cr
$ \bullet$  & Product states for the observables associated with two bounded spacelike separated space--time
regions $\O_1$, $\O_2$ exist if (see Theorem 6.1 below) and only if (see 
Theorem 6.2 below) the model has decent phase space properties.
\cr
$ \bullet$  & If a single product state exists, then very specific product states exist. In fact, for
any pair of partial states on the sub algebras
there exists a product state which can not be distinguished from the partial 
states by measurements in the sub algebras alone (see [4]
for a proof of these statements).
\cr
}  
\vskip .3cm
\noindent

\vskip .5cm
We conclude this introduction with some comments concerning
thermal equilibrium states,
the Hilbert spaces associated with them and the density matrices 
describing deviations from the thermal equilibrium state (see e.g.\ [8]):

\vskip .3cm
Just like in standard quantum mechanics, the observable quantities of a thermal theory
are modelled by linear operators which are embedded into an algebra 
${\cal R}_\beta \subset \B(\H_\beta)$ of bounded operators acting on a separable Hilbert space $\H_\beta$.
One may in fact start from an {\it abstract} algebra of observables $\A$ (which might e.g.\ be constructed from the
$C^*$-algebra associated with the Poincar\'e group (see [9])). The algebra ${\cal R}_\beta$ and
the Hilbert state $\H_\beta$ may then be viewed as secondary objects which arise via the 
GNS construction, once a thermal state $\omega_\beta$ on $\A$ has been distinguished. More precisely, 
the algebra ${\cal R}_\beta$ can be viewed as the weak closure $\pi_\beta (\A)''$ of 
$\pi_\beta (\A)$ in $\B(\H_\beta)$. The vector $\omega_\beta$ can be identified as the GNS vector 
associated with the pair $(\A, \omega_\beta)$, which satisfies
\# { \omega_\beta (a) = \bigl( \Omega_\beta \, , \, \pi_\beta (a) \Omega_\beta \bigr) \qquad 
\forall a \in \A.}
Thermal states are always mixed states, i.e., they can be decomposed into a convex
combination of states (which are not necessarily equilibrium states). 
However, if we restrict the decomposition to thermal states, then it is unique.
An equilibrium state is called extremal, if it cannot 
be decomposed into other equilibrium states. 
The decomposition into extremal equilibrium states
corresponds to the physical separation of an equilibrium state into 
pure thermodynamic phases. The symmetry, or lack of symmetry, of these phases is thereby automatically determined.
In this short letter we will concentrate on pure phases. Therefore we will always
assume that the given equilibrium state is an extremal equilibrium state.

Extremal equilibrium states are distinguished 
within the set of all (physical) states of $\A$ by first principles: they are precisely those states which are
distinguished among (possible other) stationary states by the fact that they 
turn continuously into the unperturbed states as a certain family of perturbations tends to
zero [10].  The same condition
may also be interpreted as adiabatic invariance [11]: 
Extremal equilibrium states return to their original form at the end of 
a procedure in which the dynamical law is changed by a local perturbation which is slowly
switched on and, as $t \to \infty$, slowly switched off again. 
A second important characteristic of equilibrium states is their passivity, 
which is the requirement 
that the energy of the system at time $t$ can only have increased if the Hamiltonian $H_\beta$
(which generates the time evolution in the thermal representation) depends on the
time and has returned to its initial form at time $t$ [12]. 
This condition is just the second law of thermodynamics; it fixes the sign of 
$\beta$ and means that no energy can be removed from an equilibrium state having $\beta >0$, 
just as a periodic process can extract no energy from the ground state.

As a sub algebra of $\B(\H_\beta)$ the algebra ${\cal R}_\beta$ generated by the observables of a thermal theory
acts reducibly, i.e.,  its commutant
\# { {\cal R}_\beta':= \{ A \in \B(\H_\beta) : [A, B] = 0 \quad \forall B \in {\cal R}_\beta \}
}
is unequal to $\C \cdot \1 $. (The underlying reason is that thermal states are mixed states.) In fact, 
${\cal R}_\beta'$ is as large as (more precisely, isomorphic to) ${\cal R}_\beta$. 
Consequently the structural properties of a thermofield theory
are somehow complementary to the ones known from zero temperature quantum field theory.
$\Omega_\beta \in \H_\beta$ induces an extremal equilibrium state if, and only if,
\# { {\cal R}_\beta \cap {\cal R}_\beta' = \C  \cdot \1 ; }
in this case $\Omega_\beta $ is said to induce a factor state and ${\cal R}_\beta$ is called a factor. 
Two extremal equilibrium states are
either equal or disjoint, which means that they represent different global circumstances.
If at a certain temperature there 
is one and only one equilibrium state, this state is automatically 
a factor state. 
Factor states have characteristic cluster properties which reflect the absence 
of long-range correlations, or the absence of large fluctuations for 
the values of space-averaged observables. In fact, it has been shown by the author [2] that
there is a tight relation between the infrared properties 
and the decay of spatial
correlations in any extremal equilibrium state, in complete analogy to the 
well understood case of the vacuum state. To be more precise, since the energy spectrum 
does not have a mass gap, the
correlations between two spacelike separated measurements
are bounded by some inverse power of their spatial distance. (The correlations of free massless
bosons in two dimensions saturate these bounds.)
These various points all indicate that pure phases should correspond to factor states 
and more precisely to extremal equilibrium states.

\vskip 1cm

\Hl{LOCAL THERMOFIELD THEORIES}

\noindent
In the framework of local quantum physics (see e.g.\ [13][14][15]) a thermofield theory is specified 
by a map (usually called a net)
\# { \O \to {\cal R}_\beta  (\O) , \qquad \O \subset \R^4, } 
which associates bounded space--time regions with algebras of
bounded operators (von Neumann algebras, to be precise) acting on a Hilbert space $\H_\beta$.
The Hermitian elements of ${\cal R}_\beta  (\O)$  are interpreted as 
the observables which can be measured at times and locations in $\O$. 
If $\O_1 \subset \O_2$, then there exists an embedding
\# { {\cal R}_\beta  (\O_1) \hookrightarrow {\cal R}_\beta (\O_2).} 
We assume that the net $\O \to {\cal R}_\beta (\O)$ is additive, i.e.,  if
\# { \cup_i \O_i = \O \Rightarrow \vee_i {\cal R}_\beta (\O_i) 
= {\cal R}_\beta (\O).} 
Here ${\cal R}_1 \vee {\cal R}_2$ denotes the von Neumann
algebra generated by the algebras ${\cal R}_1 $ and ${\cal R}_2$ (i.e., the smallest von Neumann
algebra containing ${\cal R}_1$ and ${\cal R}_2$).
The algebra
\# {{\cal R}_\beta  := \vee_{ {\cal O}  \subset \r^4} {\cal R}_\beta (\O)}
is defined as the inductive limit of the local algebras (see [16].).
Observables localized in spacelike separated space--time regions commute:
\#
{{\cal R}_\beta (\O_1) \subset {\cal R}_\beta  ( \O_2)' \quad \hbox{\rm if} \quad \O_1 \subset \O_2'.}
Here $\O '$ denotes the spacelike complement of $\O$ and 
${\cal R}_\beta (\O)'$ denotes the set of operators in~$\B (\H_\beta)$ which commute with all operators in ${\cal R}_\beta (\O)$.
This property is traditionally called {\sl locality}, since it reflects the local character of the interaction. However, in order to avoid 
confusion with the usage of this term in the current literature on quantum information theory, we will prefer the  
german term ``Nahwirkungsprinzip''.

As mentioned in the introduction, the decay of correlations or the absence of large fluctuations is typical of pure thermodynamic 
phases. But decent cluster properties can only be expected if the KMS state can not be decomposed into 
time invariant states. An appropriate criterion is available: 
An extremal KMS state is an extremal time invariant state,
if and only if, it is weakly asymptotically abelian, i.e.,
\# {\lim_{s-r \to \infty} \int_s^r {\rm d}t \, \, ( \Omega_\beta \, , \, A [ {\rm e}^{-it H_\beta} 
B  {\rm e}^{it H_\beta} , C ] D
\Omega_\beta ) = 0 }
for all $A,B,C, D \in {\cal R}_\beta$ [8, 2.5.31]. (We should note that the 
Hamiltonian $H_\beta$ which generates the (weakly continuous) one-parameter unitary group of
time evolutions in the thermal representation is not bounded from below for $T> 0$.) 
The property (13) excludes finite quantum systems; a purely discrete energy spectrum
leads to a quasi-periodic motion violating
the condition of weak asymptotic abelianess. For infinite quantum systems, however,
weak asymptotic abelianess can be considered as a generic feature, and if it holds, then
the algebra of observables ${\cal R}_\beta$ is of type III in the classification of Murray and von Neumann
[8, 5.3.36]. 
Consequently, it does not contain any finite dimensional projection.
(A projection $P=P^2= P^* \in \B(\H_\beta)$ is called finite dimensional if $P\H_\beta$ is a finite
dimensional subspace of $\H_\beta$.)

If the time evolution is norm asymptotically abelian, i.e.,
\# { \lim_{t \to \infty} \bigl\| [ A,  {\rm e}^{-it H_\beta} B  {\rm e}^{it H_\beta} ] \bigr\| = 0 \qquad 
\forall  A,B \in {\cal R}_\beta ,} 
then $\Omega_\beta$ is the unique---up to a phase---time invariant vector 
in $\H_\beta$. We will assume the latter property in the sequel.   

\vskip 1cm

\Hl{THE REEH--SCHLIEDER PROPERTY}

\noindent
A relativistic theory requires a drastic departure from `classical' quantum mechanics and its interpretation. 
The famous Reeh--Schlieder theorem [17] states that if there where no 
restrictions on the 
available energy--momentum transfer, then one could prepare any vector state with arbitrary accuracy using only 
strictly local operations; i.e., operations 
performed in an arbitrary bounded space--time region. 
However, we emphasize that the cluster theorems put severe limits on the size of 
affordable effects, if the available energy--momentum transfer is restricted. 

In thermofield theory the thermal equilibrium state itself selects (via the GNS construction)
an appropriate
Hilbert space $\H_\beta$. All states which represent the same global circumstances
(mean particle density, mean energy density, temperature, etc.)
can be described by density matrices acting on $\H_\beta$.
The thermal equilibrium state itself is induced by a vector $\Omega_\beta \in \H_\beta$ 
which satisfies [18] (and is completely characterized by)
the Kubo--Martin--Schwinger (KMS) boundary condition [19][20]
\# { (\Omega_\beta \, , \, A {\rm e}^{- \beta H_\beta} B \Omega_\beta )= (\Omega_\beta \, , \, BA \Omega_\beta ) \qquad \forall A, B \in {\cal R}_\beta .}
The KMS vector $\Omega_\beta$ (sometimes called the thermal vacuum vector)
is cyclic, i.e.,
\# { \overline { {\cal R}_\beta \Omega_\beta } = \H_\beta }
and separating for ${\cal R}_\beta$, i.e.,
\# { A \Omega_\beta  = 0 \quad \Rightarrow \quad A= 0 \qquad \forall A \in {\cal R}_\beta.   }
Thus any state, which is described by a density matrix $\rho \in \B(\H_\beta)$ 
(i.e., any state describing the same global circumstances as the thermal state w.r.t.\
mean particle density, mean energy density, temperature, etc.)
is a vector state (see e.g.\ [8][16]).
As we will discuss next, $\Omega_\beta$ is even cyclic for ${\cal R}_\beta (\O)$, if $\O$ contains an open
space--time region $\O_\circ$. 
In order to derive this result, we have to take a closer look at the characteristic analyticity properties of 
a relativistic KMS states.

{\it Lorentz invariance} is always broken by a KMS state [21][22].  
A KMS state might also break {\it spatial translation} or 
{\it rotation invariance}, but
the maximal propagation velocity of signals, which is characteristic for a relativistic theory,
is not affected by such
a lack of symmetry. 
It was first recognized by Bros and Buchholz that a finite maximal propagation velocity
of signals implies that the KMS states of a {\sl relativistic} theory have stronger 
analyticity properties in configuration space than those imposed by the traditional 
KMS condition~[23].

\Def{A vector  $\Omega_\beta $ satisfies the {\it relativistic KMS condition}   
at inverse temperature $\beta > 0$ if and only
if there exists some positive timelike vector $e \in V_+$, $e^2 = 1$, such that 
for every pair of elements $A, B$ of ${\cal R}_\beta$ there exists a function~$F_{A,B}$ which is
analytic in the domain
\# { -{\cal T}_{\beta e /2} \times {\cal T}_{\beta e /2},}
where ${\cal T}_{\beta e /2} = \bigl\{ z \in \C : \Im z \in V_+ \cap ( \beta e / 2 + V_- ) \bigr\}$ is a
tube, and  continuous at the boundary sets 
$\R^4 \times \R^4$ and $\bigl(\R^4 - {i \over 2} \beta e \bigr) \times \bigl(\R^4 
+ {i \over 2} \beta e \bigr)$ with boundary values given by
\& 
{ F_{A,B} (x_1, x_2) & = \bigl( \Omega_\beta \, , \,   \alpha_{x_1} ( A) \alpha_{x_2} (B) \Omega_\beta \bigr)  
\cr
F_{A,B} \Bigl( x_1 - {i \over 2} \beta e, 
x_2 + {i \over 2} \beta e \Bigr) & = \bigl( \Omega_\beta \, , \, \alpha_{x_2} (B)
\alpha_{x_1} (A) \Omega_\beta \bigr)   \qquad \forall x_1, x_2 \in \R^4 .}  
Here $\alpha_x (A)$ means that the element $A \in {\cal R}_\beta$ has been shifted in space--time by
$x \in \R^4$. }

\Rem{The relativistic KMS condition can be understood as a remnant of the relativistic 
spectrum condition in the vacuum sector. It has been rigorously established (see [23])
for the KMS states constructed by Buchholz and Junglas [24]. 
We would like to emphasize that the relativistic KMS condition {$\underline {\hbox{does  not}}$} 
exclude the possibility that
the thermal state breaks translation or rotation symmetry. (Consequently, a unitary implementation
of the spatial translations may not exist. This is why we had introduce the automorphisms
$\alpha_x \colon {\cal R}_\beta (\O) \to {\cal R}_\beta (\O+x)$, $\O \subset \R^4$, in the previous definition.)}

As has been demonstrated by the author [5, Th.\ 3.9], the relativistic KMS condition implies the 
Reeh--Schlieder property:
$\Omega_\beta$ is cyclic for ${\cal R}_\beta (\O)$, i.e., 
\# { \overline { {\cal R}_\beta (\O) \Omega_\beta} = \H_\beta, }
where $\O$ is any open subset of~$\R^4$. 

We can reformulate this result:

\Th{Given a density matrix $\rho \in \B(\H_\beta)$ and any $\epsilon > 0$
we can find an element $A_{\rho, \epsilon} \in {\cal R}_\beta (\O)$ 
(representing a strictly local operation in $\O$)
such that 
\# { \sup_{\| B \| =1, B \in {\cal R}_\beta} \Bigl| \Tr \, \rho B - (A_{\rho, \epsilon} \Omega_\beta \, , \, B 
A_{\rho, \epsilon} \Omega_\beta ) \Bigr| < \epsilon.}
}

\vskip -.6cm
\Rem{As mentioned before, any state which is described by a density
matrix is a vector state. The theorem given is an immediate consequence of (20) and this fact.
It has been shown in [5, Th.\ 3.10] that $\Omega_\beta$  
shares the ``Reeh-Schlieder property'' (20)  with a dense set of 
vectors in $\H_\beta$.}

As has been shown by Kadison [25], the Reeh--Schlieder property implies that the local algebras
can not be finite dimensional matrix algebras:

\Prop{(Kadison).
If ${\cal R}_\circ$ is a proper sub von Neumann algebra of the von Neumann algebra ${\cal R}$,
and $\Omega$ is a separating and cyclic vector for both ${\cal R}$ and ${\cal R}_\circ$, then 
${\cal R}$ (and ${\cal R}_\circ$) are of infinite type.}

\vskip 1cm

\Hl{THE SCHLIEDER AND THE BORCHERS PROPERTY}

\noindent
Without any further assumptions on e.g., the energy spectrum of excitations
of a given extremal KMS state, the following statement is valid:

\Th{\hbox{\rm (Schlieder property).}
Let $\O$ and $\hat {\O}$ denote two open (not necessarily bounded) 
space--time regions such that 
\# { \O + t e \subset \hat {\cal O}  
\qquad \forall |t| < \delta , \quad \delta >0.}
It follows that $0 \ne A \in {\cal R}_\beta (\O)$ and 
$0 \ne B \in {\cal R}_\beta (\hat {\O})'$
implies $AB \ne 0$.}

\Rem{A proof of the thermal case given here can be found in [3, Th.\ II.4]; the original
result for the vacuum sector can be found in [26].}

Thus, if $P \in {\cal R}_\beta (\O)$ and 
$Q \in {\cal R}_\beta (\hat {\O}') \subset {\cal R}_\beta (\hat {\O})' $ are two nontrivial
projections, then the product PQ can not vanish identically. 
The Schlieder property implies that ${\cal R}_\beta (\O)$ is almost a factor, namely
\# { {\cal R}_\beta (\O) \cap {\cal R}_\beta (\hat{\O})' = \C \cdot \1 .}
It is a first step towards
the ``statistical independence'' of ${\cal R}_\beta (\O)$ and
${\cal R}_\beta (\hat{\O})'$: 

\Cor{\hbox{\rm (Florig and Summers [27])}: Let $\O$, $\hat{\O}$ denote 
a pair of space--time regions such that the closure of the open 
region $\O$ is contained in the interior of $\hat{\O}$.
It follows that
\vskip .3cm
\noindent
\halign{ #  \hfil & \vtop { \parindent =0pt \hsize=36,6em
                            \strut # \strut} \cr 
(i)   &  For any nonzero vectors $\Phi$, $\Psi \in \H_\beta$ there 
exist $A' \in {\cal R}_\beta (\O)' $ and 
$B' \in {\cal R}_\beta (\hat{\O})$ such 
that  
\# { A'\Phi = B'\Psi \ne 0.}
\vskip -.2cm
\cr
(ii)   &  $\| A B \| = \| A \| \,  \| B \|$
for all $A \in {\cal R}_\beta (\O)$ and all $B \in {\cal R}_\beta (\hat{\O})'$.
\cr
(iii)   &  The von Neumann algebras ${\cal R}_\beta (\O)$ and 
${\cal R}_\beta (\hat{\O})'$ 
are algebraically independent; i.e.,  
given  two arbitrary sets $\{ A_i : i = 1, \ldots, m \} $ 
and $\{ B_j : j = 1, \ldots, n \} $ 
of linear independent elements of ${\cal R}_\beta (\O)$ 
and ${\cal R}_\beta (\hat{\O})'$, 
respectively, 
the collection $\{ A_i B_j : i = 1, \ldots, m ; j = 1, \ldots, n \}$ 
is linearly 
independent in the algebraic tensor product 
${\cal R}_\beta (\O) \odot {\cal R}_\beta (\hat{\O})'$
of ${\cal R}_\beta (\O)$ and ${\cal R}_\beta (\hat{\O})'$.
\cr
}  
}

The Schlieder property has interesting consequences for the 
existence of certain states. E.g., for any pair of
nontrivial projections (i.e., YES-NO experiment)
$E \in {\cal R}_\beta (\O)$, $F \in {\cal R}_\beta (\hat{\O}')$ 
and $\lambda, \mu \in [0, 1]$ 
there exists a density matrix $\rho \in \B(\H_\beta)$,
such that 
\# { \Tr \, \rho E = \lambda \quad \hbox{and}  \quad \Tr \, \rho F = \mu.} 
Moreover, for any pair of density matrices $\rho_1, \rho_2  \in \B(\H_\beta)$
there exists a density matrix $\rho \in \B(\H_\beta)$ such that
\# {\Tr \, \rho_1 A = \Tr \, \rho A  \qquad \forall A \in {\cal R}_\beta (\O)}
and 
\# {\Tr \, \rho_2 B = \Tr \, \rho B  \qquad \forall B \in {\cal R}_\beta  (\hat{\O}').} 
A proof of these statements can be found in [27].

\bigskip
Another consequence of the ``Nahwirkungsprinzip'', the Reeh--Schlieder property
and the 
Schlieder property is the so-called Borchers property:

\Th{\hbox{\rm (Borchers property).} Let $\O$ and $\hat {\O}$ denote two open and bounded
space--time regions such that 
\# { \O + t e \subset \hat {\cal O}  
\qquad \forall |t| < \delta , \quad \delta >0.}
Given a nonzero projection $E \in {\cal R}_\beta (\O)$, there exists a partial
isometry $V \in {\cal R}_\beta (\hat {\cal O})$ such that $V^* V = \1 $ and $V V^* = E$.}

\Rem{One writes  
\# { E \sim \1 _{\hbox{\fiverm mod} \, {\cal R}_\beta (\hat {\cal O}) } \, . }
Recall that a factor ${\cal M}$ is called type~III, if 
$E \sim \1 _{\hbox{\fiverm mod} \, {\cal M} } $ for all self-adjoint
projections $E \in {\cal M}$. Thus ${\cal R}_\beta (\O)$ is of infinite type, and ``almost''
a factor of type~III.
A proof of the thermal case given here can be found in [3, Th.\ II.6]; the original
result for the vacuum sector can be found in [28].}

The Borchers property has interesting consequences for the actual preparation of 
states: Given a density matrix $\rho$, we set
$ \rho_V  := V \rho V^* $.
Then
\# { \Tr  \, \rho_V E = 1  \qquad \hbox{and} 
\qquad \Tr \, \rho_V B  = \Tr \, \rho B \quad 
\forall B \in {\cal R} (\hat {\cal O}').}
This demonstrates that the Borchers property allows us to prepare a state $\omega_V$
which satisfies the properties (28) by a strictly local operation. 
The state given remains
completely unchanged in the  spatial complement of $\hat {\O}$. This is a remarkable 
difference to the ``collapse of the wave-function'' type of preparation. 

\Rem{While finite quantum systems may very well allow a tensor product decomposition of their observables into
sub algebras, which are statistically independent (so that they satisfy the Schlieder and even
the split property (see Sec.\ 6)), the Borchers property 
is reserved for infinite quantum systems (as can be seen from the previous remark).
Thinking of photon polarization states, which are entangled into a Bell state $\rho$ on $\C^2 \otimes \C^2$, 
it will not always be possible to find an isometry $V \in \B(\C^2 \otimes \C^2)$ 
(isometries are unitary operators in the finite dimensional case)
such that for 
some given projection $E \otimes \1 $ 
the resulting state $\rho_V$
satisfies the  properties 
\# { \Tr  \, \rho_V E = 1  \qquad \hbox{and} 
\qquad \Tr \, \rho_V B  = \Tr \, \rho B \quad 
\forall B \in \1 \otimes \B(\C^2).}
The equation on the r.h.s.\ implies that $V \in \B(\C^2) \otimes \1 $. If, for example the restriction of $\rho$
to $\B(\C^2) \otimes \1 $ coincides with ${1 \over 2} E \otimes \1 + {1 \over 2} (\1 - E) \otimes \1 $
then one can not find a unitary operator in $\B(\C^2) \otimes \1 $ such that the equation on the l.h.s.\
in (31) is fulfilled.
The obvious difference is that in the case discussed in
Theorem 4.3 an intermediate region $\hat \O \setminus \O$ is at our free disposal, which may be used to 
decouple the inside from the outside.}

\vskip 1cm

\Hl{ENTANGLED STATES}

\noindent
\Def{Consider two sub algebras ${\cal R}_1$ and ${\cal R}_2$ of a von Neumann algebra ${\cal R} \subset \B(\H)$. 
A state induced by a density matrix $\rho \in \B(\H)$ on ${\cal R}$ is 
called a {\it product state} for the pair $({\cal R}_1, {\cal R}_2)$ if there exist two density matrices 
$\rho_1, \rho_2 \in \B(\H)$  such that
\# { \Tr \,  \rho AB  = \Tr \, \rho_1 A  \cdot  \Tr \, \rho_2 B }
for all $A \in {\cal R}_1$ and $B \in {\cal R}_2$.  A state induced by a density matrix $\rho \in \B(\H)$
is called {\it separable}
for the pair $({\cal R}_1, {\cal R}_2)$, if, and only if, it is in the weak$^*$-closed convex 
hull of product states.
I.e., there exists a weak$^*$-limit of convex combinations of
product states $\{ \rho_\gamma \}_{\gamma \in I}$ for the pair $({\cal R}_1, {\cal R}_2)$ such that  
\# { \Tr \, \rho AB  = \sum_{\gamma \in I} \lambda_\gamma \bigl( \Tr \,  \rho_\gamma AB \bigr), 
\qquad \sum_{\gamma \in I} \lambda_\gamma =1, \quad 0 < \lambda_\gamma < 1, 
}
for all $A \in {\cal R}_1$ and $B \in {\cal R}_2$. A density matrix $\rho \in \B(\H)$ is 
called {\sl entangled} w.r.t.\ the pair $({\cal R}_1, {\cal R}_2)$, if it is not separable. }

In a recent paper Clifton and Halvorson [1] derived the following result:

\Prop{\hbox{\rm (Clifton and Halvorson).}
Let ${\cal R}_1$ and ${\cal R}_2$ be nonabelian von Neumann algebras acting on a Hilbert space $\H$ such that
${\cal R}_1 \subset {\cal R}_2'$ and ${\cal R}_1$, ${\cal R}_2$ satisfy the Schlieder property. If
${\cal R}_1$ and ${\cal R}_2$ are of infinity type, then there is an open dense subset of vectors in $\H$,
which induce entangled states w.r.t.\ the pair $({\cal R}_1, {\cal R}_2)$.
If a vector  $\Phi\in \H$ is cyclic for ${\cal R}_1$, then $(\Phi \, , \, . \, \Phi)$ is 
one of these entangled states w.r.t.\ the pair $({\cal R}_1, {\cal R}_2)$.}

As mentioned before, any state which is described by a density matrix acting on $\H_\beta$ is a vector state.
Moreover, $\Omega_\beta$ shares the Reeh--Schlieder property with a dense set of vectors. 
Thus the following corollary is an immediate consequence of Proposition 5.1.

\Cor{Let $\O_1$ and $\O_2$ denote two open (not necessarily bounded) 
space--time regions such that 
\# { \O_1 + t e \subset \O_2'  
\qquad \forall |t| < \delta , \quad \delta >0.}
Then there exists a norm dense set of entangled states w.r.t.\ the pair $\bigl({\cal R}_\beta (\O_1),  {\cal R}_\beta (\O_2)\bigr)$.}

\Pr{As mentioned before, Kadison's result (see Prop.\ 3.2) and the Reeh--Schlieder
theorem together imply that ${\cal R}_\beta (\O_1)$ and ${\cal R}_\beta (\O_2)$ are of infinite type.
There are now two distinct lines of arguments which may be used to prove the Corollary given: 
\vskip .2cm
\noindent
i.) According to Prop.\ 2 of [1], if $\Phi$ is a cyclic vector for ${\cal R}_1$, then $\Phi$ is entangled 
w.r.t.\ the pair $({\cal R}_1, {\cal R}_2)$ (as long as ${\cal R}_1$ and ${\cal R}_2$ are nonabelian.)
Consequently, the state induced by $\Omega_\beta$ is entangled. Moreover,  
there is a dense set of vectors in $\H_\beta$ which share the Reeh--Schlieder property with $\Omega_\beta$.
These vectors induce entangled states w.r.t.\ the pair 
$\bigl({\cal R}_\beta (\O_1), {\cal R}_\beta (\O_2) \bigr)$.
\vskip .2cm
\noindent
ii.) According to Prop.\ 1 of [1],  if ${\cal R}_1$ and ${\cal R}_2$ satisfy the Schlieder property
and are both  of infinite type, 
then there is a dense set of Bell correlated (hence entangled) states 
w.r.t.\ the pair $({\cal R}_1, {\cal R}_2)$. This result is in one sense 
stronger then the first, since a mixed entangled state need not display Bell correlations.
But this second line of arguments fails to establish that the vector state induced by
$\Omega_\beta$ (or any other experimentally feasible vector state is entangled.}

This result may be compared with the finite dimensional situation: If ${\cal R}_1 := \B(\C^2) \otimes {\1 }_2$,
${\cal R}_2 :={\1 }_2 \otimes \B(\C^2)$, then any entangled state vector is cyclic for ${\cal R}_1$.
But the set of entangled states of $\B(\C^2) \otimes  \B(\C^2)$ is not norm dense in the set of all density matrices on
$\B(\C^2) \otimes  \B(\C^2)$ [29]. 

\vskip 1cm

\Hl{THE SPLIT PROPERTY}

\noindent
While the existence of a norm dense set of entangled states is a direct consequence of the generic
properties of any thermofield theory, the existence of (normal) product states is a much more delicate
property. 
If one carefully studies (Sect.\ 3 in [4] is entirely devoted to this task)
the distribution of the energy eigenvalues of a system consisting of
relativistic particles  which  are confined in a finite volume  
as the size of the ``box'' goes to infinity,
then one is led to the assumption that a thermal state
should have the following characteristic phasespace properties: for any bounded space--time region $\O$
the maps $\Theta^+_{\alpha, {\cal O}} \colon {\cal R}_\beta (\O) \to \H_\beta $ given by 
\# { \Theta^+_{\alpha, {\cal O}} (A) = {\rm e}^{ -\alpha H_\beta} P^+ A \Omega_\beta , \qquad \alpha > 0, }
are of type $s$. 
Here $P^+$ denotes the 
projection onto the strictly positive spectrum of the Hamiltonian $H_\beta$. 
We recall (see e.g.\ [30]) that  
$\Theta^+_{\alpha, {\cal O}}$ is said to be of 
type~$l^p$, $p > 0$, if there exists
a sequence of linear mappings $\Theta_k$ of rank $k$ such that
\# { \sum_{k=0}^\infty \| \Theta^+_{\alpha, {\cal O}} - \Theta_k \|^p < \infty  .}  
$\Theta^+_{\alpha, {\cal O}}$ is said to be nuclear, if 
$\Theta^+_{\alpha, {\cal O}}$ is of type~$l^p$ for $p =1$. 
It is said to be of type  $s$, if 
it is of type~$l^p$ for all $p > 0$. 
Quantitative information can be extracted from the nuclear norm of $\Theta^+_{\alpha, {\cal O}}$, but
we will not need this kind of information here.

\Th{\hbox{\rm (Split Property).} Assume that for any bounded space--time region $\O$ 
the maps $\Theta^+_{\alpha, {\cal O}}$
are of type $s$ for all $\alpha > 0$. 
It follows that for any pair $\Lambda = (\O,\hat{\O})$  of open bounded space--time regions 
which satisfies
\# { \O + t e \subset \hat{\O}  
\qquad \forall |t| < \delta , \quad \delta >0,}
\vskip .3cm
\noindent
\halign{ #  \hfil & \vtop { \parindent =0pt \hsize=36,6em
                            \strut # \strut} \cr 
(i) & there exists a  vector $\eta_\Lambda \in \H_\beta$ 
such that
\vskip .1cm
\hskip .5cm
a.) $(\eta_\Lambda , A B \eta_\Lambda) 
= (\Omega_\beta , A \Omega_\beta)  (\Omega_\beta, B \Omega_\beta)$
for all $A \in {\cal R}_\beta (\O)$ and
$B \in {\cal R}_\beta (\hat {\cal O})'$.
\vskip .1cm
\hskip .5cm
b.) $\eta_\Lambda$ is cyclic and separating for ${\cal R}_\beta (\O) 
\vee {\cal R}_\beta ( \hat {\cal O})'$.
\vskip .2cm
\cr
(ii)    & the von Neumann algebra generated by ${\cal R}_\beta (\O) $
and ${\cal R}_\beta (\hat {\cal O})' $ is isomorphic to the 
$W^*$-tensor product of the two algebras.
I.e.,  there exists a unitary operator 
$W_\Lambda \colon \H_\beta \to \H_\beta \otimes \H_\beta$  such that
\# { W_\Lambda A B W_\Lambda^* = A \otimes B } 
for all $A \in {\cal R}_\beta (\O)$ and
$B \in {\cal R}_\beta (\hat {\cal O})'$.
\cr
}  
}

\Rem{This result has been proven by the author in [4, Sect.\ 4]; the proof is based
on previous work by different authors, see for instance [31]. We emphasize, that the the requirement 
that $\O$ is bounded can not be removed (see e.g.\ [32] for counterexamples, which may easily be adapted to the
thermal case).}

The split property (see [33] for a general account) 
has far reaching consequences for the preparation of states:
It implies that for any pair of density matrices $\rho_1 , \rho_2 \in \B(\H_\beta)$
there exists a density matrix $\rho \in \B(\H_\beta)$ such that
\# {\Tr \, \rho \, AB  = \Tr \, \rho_1 A \cdot \Tr \, \rho_2 B}
for all $A \in {\cal R}_\beta (\O)$ and $B \in {\cal R}_\beta (\hat {\cal O})'$.
Moreover, it allows us to select a set of states which represents local excitations of the thermal
states: the set 
\# {  {\cal L}_\beta (\O, \hat{\O}) := \overline{ {\cal R}_\beta (\O) \eta_\Lambda  } }
has the following properties:
\vskip .3cm
\halign{ \indent #  \hfil & \vtop { \parindent = 0pt \hsize=34em
                            \strut # \strut} 
\cr 
(i)    & ${\cal L}_\beta (\O, \hat{\O})$ is 
a closed subspace of $\H_\beta$;
\cr
(ii)    & ${\cal L}_\beta (\O, \hat{\O})$ is invariant under the action of 
${\cal R}_\beta (\O)$;
\cr
(iii)    & If $\Psi \in {\cal L}_\beta (\O, \hat{\O})$, then  
\# {(\Psi \, , \, AB \Psi) = (\Psi\, , \, A \Psi)  ( \Omega_\beta \, , \, B \Omega_\beta) }
for all  
$A \in {\cal R}_\beta (\O)$ and $B \in {\cal R}_\beta (\hat{\O})'$.
\cr
(iv)    & ${\cal L}_\beta (\O, \hat{\O})$ is complete in the following sense: 
to every density matrix $\rho \in \B(\H_\beta)$
there exists a vector $\Phi \in {\cal L}_\beta (\O, \hat{\O})$ such that
\# {(\Phi\, , \, A \Phi) = \Tr \, \rho A  }
for all $A \in {\cal R}_\beta (\O)$. 
\cr
}  

\Rem{Similar properties have been derived for a subspace of the vacuum Hilbert space representing 
local excitations of the vacuum (see [24]). The generalization to the present thermal
case is straight forward.}

It was noticed by Buchholz, D'Antoni and Longo
that the split property imposes certain restrictions on the energy level density
of excitations of the KMS state [31 , Proposition 4.2 and Lemma 3.1.i).]:

\Th{\hbox{\rm (Buchholz, D'Antoni and Longo)}. Consider a TFT, specified by a
von Neumann algebra ${\cal R}_\beta$ with a cyclic and separating
vector $\Omega_\beta$ and a net of sub algebras
$\O \to {\cal R}_\beta (\O)$,  
subject to the conditions specified in Sect.\ 2.
Assume the inclusion 
${\cal R}_\beta (\O) \subset {\cal R}_\beta (\hat{\O})$
is split. Then the maps
\# { \Theta^+_{\alpha, {\cal O}} (A) = {\rm e}^{ -\alpha H_\beta} P^+ A \Omega_\beta , \qquad \alpha > 0, }
are compact for $\alpha > 0$. } 

\Rem{Let us summarize: If the map $\Theta^+_{\alpha, {\cal O}} $ is of type $s$, then the split property holds. 
The latter implies that $\Theta^+_{\alpha, {\cal O}} $ is a least compact. Whether or not the
first assumption can be relaxed and whether or not the second conclusion can be sharpened is unknown to the author.}
\vskip 1cm

\noindent
{\it  Acknowledgments.\/}
\noindent
A number of helpful suggestions and remarks by the referees and Rainer Verch (Univ.\ G\"ottingen)
are gratefully acknowledged. This work was supported 
by the Fond zur F\"orderung der Wissenschaft\-lichen Forschung in Austria, Proj.\ Nr.\  Z30 THP. 
The final version of this letter was prepared at the ESI (Erwin Schr\"odinger Institute) Vienna. 

\vskip 1cm

\noindent
{\fourteenrm References}
\nobreak
\vskip .3cm
\nobreak
\halign{   &  \vtop { \parindent=0pt \hsize=33em
                            \strut  # \strut} \cr 
\HEP
{1}
{H.\ Halvorson and R.\ Clifton}  {Generic Bell correlations between arbitrary local algebras in quantum field theory}
                                  {math-ph/9909013} 
\REF
{2}
{C.\ J\"akel}                   {Decay of spatial correlations in thermal states}
                                  {Ann.\ l'Inst.\ H.\ Poincar\'e} 
                          {69}	{425--440}
                          {1998}
\REF
{3}
{C.\ J\"akel}                   {Two algebraic properties of thermal quantum field theories}
                                  {\JMP} 
                          {40}	{6234--6244}
                          {1999}
\HEP
{4}
{C.\ J\"akel}                   {Nuclearity and split for thermal quantum field theories}
                                  {hep-th/9811227} 
\REF
{5}
{C.\ J\"akel}                    {The Reeh--Schlieder property for 
                                    thermal field theories} {\JMP }
                                   {41} {1745--1754}
                                   {2000}
\REF
{6}
{J.S.\ Summers and R.F.\ Werner} {On Bell's inequalities and quantum field theory. II.
                                   Bell's inequalities are maximally violated in the vacuum} 
                                  {\JMP} 
                                  {28}	{2440--2447}
                                  {1987}
\REF
{7}
{L.\ Landau}     {On the violation of Bell's inequality in quantum theory}
                                   {Phys.\ Lett.} 
                                                      {120} {54--56}
						                                                {1987}
\BOOK
{8}  
{O.\ Bratteli and D.W.\ Robinson} {Operator Algebras and Quantum Statistical Mechanics~I,II} 
                                  {Sprin\-ger-Verlag, New York-Heidelberg-Berlin} 
                                  {1981}
\BOOK
{9}
{W.\ Kunhardt}    {\"Uber die universelle Algebra der lokalen Translationsoperatoren in der Algebraischen Quantenfeldtheorie}    
              {Diplomarbeit, Universit\"at Hamburg}                                       
              {1994}                                       
\REF
{10}
{R.\ Haag, D.\ Kastler, and E.B.\ Trych-Pohlmeyer}    {Stability and equilibrium states}
                                                      {\CMP} 
                                                      {38} {173--193}
						                                                {1974}
\REF
{11}
{H.\ Narnhofer and W.\ Thirring}   {Adiabatic Theorem in Quantum Statistical Mechanics}
                                   {Phys.\ Rev}
                                   {A26/6}   {3646--3652} 
                                   {1982}
\REF
{12}
{W.\ Pusz and S.L.\ Woronowicz}   {Passive states and KMS states for general 
               quantum systems}
              {\CMP}
             {58}    {273--290}
             {1978}
\BOOK
{13}  
{H.\ Araki} {Mathematical Theory of Quantum Fields} 
                                  {Oxford University Press, Oxford} 
                                  {1999}
\BOOK
{14}
{R.\ Haag}    {Local Quantum Physics: Fields, Particles, Algebras} 
              {Springer-Verlag, Berlin-Heidel\-berg-New York} 
              {1992}
\REF
{15}
{R.\ Haag and D.\ Kastler}        {An algebraic approach to quantum field theory}
                                   {\JMP}
                                   {5}  {848--861}
                                   {1964}
\BOOK
{16}
{R.V.\ Kadison and J.R.\ Ringrose}     {Fundamentals of the theory of operator algebras II} 
                   {Academic Press, New York}  
                  {1986}
\REF
{17}
{H.\ Reeh and S.\ Schlieder}     {Bemerkungen zur Unit\"ar\"aquivalenz von 
                                    Lorentzinvarianten Feldern}
                                   {Nuovo Cimemento} 
                                                      {22} {1051--1056}
						                                                {1961}
\REF
{18}
{R.\ Haag, N.M.\ Hugenholtz, and M.\ Winnink}
                          {On the equilibrium states in quantum statistical mechanics}  
                          {\CMP}	
                          {5}	{215--236}
                          {1967}
\REF
{19}
{R.\ Kubo}    {Statistical mechanical theory of irreversible processes I.}    
              {J.\ Math.\ Soc.\ Jap}                                       
              {12} {570--586}                                       
              {1957}                                       
\REF
{20}
{P.\ C.\ Martin and J.\ Schwinger}   {Theory of many-particle systems.\ I}
                                    {Phys.\ Rev}
                                    {115/6} {1342--1373}
                                    {1959}
\REF
{21}
{H.\ Narnhofer}   {Kommutative Automorphismen und Gleichgewichtszust\"ande}
                  {Act.\ Phys.\ Austriaca}
                  {47}   {1--29} 
                  {1977}
\REF
{22}
{I.\ Ojima}   {Lorentz invariance vs. temperature in QFT}
              {Lett.\ Math.\ Phys}
              {11}    {73--80}
              {1986}
\REF
{23}
{J.\ Bros and D.\ Buchholz}      {Towards a relativistic KMS condition}
                                  {Nucl.\ Phys.\ B} 
                                  {429} {291--318}
                                  {1994}
\REF
{24}
{D.\ Buchholz and P.\ Junglas}   {On the existence of equilibrium states in local 
                                   quantum field theory} 
                                  {\CMP} 
                                  {121} {255--270}
                                  {1989}
\REF
{25}
{R.\ Kadison}    {Remarks on the type of von Neumann algebras of local observables in quantum field
                  theory}    
              {\JMP}                                       
              {4} {1511--1516}                                       
              {1963}                                       
\REF
{26}
{S.\ Schlieder}               {Einige Bemerkungen \"uber Projektionsoperatoren}
                                    {\CMP}
                                    {13} {216--225}
                                   {1969}
\REF
{27}
{M.\ Florig and S.J.\ Summers}      {On the statistical independence of algebras of observables}
                       {\JMP}
                       {38/3} {1318--1328}
                       {1997}
\REF
{28}
{H-J.\ Borchers}     {A remark on a theorem of B.\ Misra}
                    	{\CMP} 
                     {4}   {315--223}
                     {1967}
\REF
{29}
{K.\ Zyczkowski, P.\ Horodecki, A.\ Sanpera, and M.\ Lewenstein}
                 {Volume of the set of separable states}
              {Phys.\ Rev.\ A}
              {58}    {883--892}
              {1998}
\BOOK
{30}
{A.\ Pietsch}     {Nuclear Locally Convex Spaces} 
                  {Springer-Verlag, Berlin-Heidelberg-New York}  
                  {1972}
\REF
{31}
{D.\ Buchholz, C.\ D'Antoni, and R.\ Longo}
                                    {Nuclear maps and modular structure II:
                                     Applications to quantum field theory}
                                    {\CMP}
                                    {129}  {115--138} 
                                    {1990}
\REF
{32}
{D.\ Buchholz}       {Product states for local algebras}
                    	{\CMP} 
                     {36}   {287--304}
                     {1974}
\REF
{33}
{S.\ Doplicher and R.\ Longo}      {Standard and split inclusions of von Neumann algebras}
                                   {Invent.\ Math.}
                                   {73}    {493--536}
                                   {1984}
\cr}

\bye